\documentclass[aps,prb,twocolumn,floats,amsmath,amssymb,superscriptaddress]{revtex4-2}
\usepackage[T1]{fontenc}
\usepackage[utf8]{inputenc}
\usepackage{graphicx}
\usepackage{epsfig}
\usepackage{amsfonts}
\usepackage{amsmath}
\usepackage{natbib}
\usepackage{multirow}
\usepackage{bm}
\usepackage{epstopdf}
\usepackage{ulem}
\usepackage{color}
\usepackage{hyperref} 
\usepackage{tablefootnote}
\usepackage{float}
\usepackage{textalpha}
\usepackage{marvosym}

\definecolor{/green}{rgb}{0.0, 0.72, 0.92}

\begin{document}
\title{Switchable polarization in non-ferroelectric SrTiO$_{3}$}
\author{Wahib Aggoune}
\email{aggoune@fhi-berlin.mpg.de}
\affiliation{The NOMAD Laboratory at Fritz-Haber-Institut der Max-Planck-Gesellschaft, Faradayweg 4-6, 14195, Berlin, Germany}
\affiliation{Institut f\"ur Physik and CSMB, Humboldt-Universit\"at zu Berlin, Berlin, Germany}

\author{Matthias Schefﬂer}
\affiliation{The NOMAD Laboratory at Fritz-Haber-Institut der Max-Planck-Gesellschaft, Faradayweg 4-6, 14195, Berlin, Germany}

\begin{abstract}

Perovskites with tunable and switchable polarization hold immense promise for unlocking novel functionalities. Using density-functional theory, we reveal that intrinsic defects can induce, enhance, and control polarization in non-ferroelectric perovskites, with SrTiO$_{3}$ as our model system. At high defect concentrations, these systems exhibit strong spontaneous polarization – comparable to that of conventional ferroelectrics. Crucially, this polarization is switchable, enabled by the inherent symmetry-equivalence of defect sites in SrTiO$_{3}$. Strikingly, polarization switching not only reverses the polarization direction and modulates its magnitude but also modifies the spatial distribution of localized defect states. This dynamic behavior points to unprecedented responses to external stimuli, opening new avenues for defect-engineered materials design.

\end{abstract}
\maketitle

\section*{INTRODUCTION}

Ferroelectrics have garnered significant attention over the past hundred years due to their fundamental materials properties and potential technological applications~\cite{rev+20nat}. These materials possess a net spontaneous polarization arising from the collective alignment of electric dipoles. Interestingly, the polarization direction can be reversibly switched by applying an external electric field, resulting in a characteristic polarization ($P$) hysteresis loop. This property is critical for memory technologies such as ferroelectric field-effect transistors used in nonvolatile memory devices~\cite{Chai+20nc,Wu+19aem}. Conventional ferroelectrics are predominantly perovskite-type oxides with the general formula $A{_x}B{_y}$O$_{3}$ ($A$ and $B$ represent cations; O is the oxygen anion; $x$,$y$=1)~\cite{Sooho+24sa}. The formation of intrinsic defects has been proposed not only to enhance existing polarization~\cite{Dong+22AS} but also to induce polarization in otherwise paraelectric perovskites~\cite{Choi+09prl,Lee+15Sc,Klyukin+17prb}. Recent progress in growth techniques, enabling precise control over sample stoichiometry, has opened new avenues for defect engineering. This, in turn, has expanded opportunities for discovering novel ferroelectric materials with unprecedented properties.

As such, recent experimental works~\cite{Lee+15Sc,Yang+15apl,Mikheev+15sr,Baki+21sr} have explored the properties of off-stoichiometric ($x$,$y$=close to 1) SrTiO$_{3}$ thin films grown under Ti-\textit{rich} and Sr-\textit{poor} conditions. This material is one of the most studied oxide perovskites to develop original ideas/concepts due to its various, fascinating physical properties~\cite{Pai+18rpp}. For the above purpose, its stable cubic phase at room temperature eases to study and disentangle the role of defects and interfacial strain in ultrathin films. Emergence of room-temperature
ferroelectricity at reduced dimensions in strain-free off-stoichiometric SrTiO$_{3}$ (STO hereafter), with a clear $P$-hysteresis, is observed~\cite{Lee+15Sc}. The polarization switching ($PS$) is demonstrated and attributed to defects~\cite{Lee+15Sc,Yang+15apl}. However, the details of the mechanism are hitherto unclear. Under similar off-stoichiometry (intrinsic defects), fascinating experiments also observed a resistive switching ($RS$) from a high- to a low-resistive state upon applied voltages, which is associated with the latter $PS$~\cite{Mikheev+15sr,Baki+21sr}. Interestingly, $RS$ varies with the defect concentration as well as from low- to room-temperature~\cite{Baki+21sr}. In the latter, STO is about 20$\%$ deficient in Sr and 10$\%$ richer in Ti atoms. This suggests a defect complex, which consists of Ti-interstitial (Ti$_{i}$) and 2$\times$Sr-vacancy ($V_{\mathrm{Sr}}$). Note that in the literature the Ti$_{i}$+$V_{\mathrm{Sr}}$ pair is called as Ti$_{\mathrm{Sr}}$ antisite~\cite{Choi+09prl,Klyukin+17prb,Lee+15Sc,Baki+21sr} although the Ti atom shifts from the Sr site. Here, from now on, we will also adopt this nomenclature of Ti$_{\mathrm{Sr}}$ antisite for the (Ti$_{i}$+$V_{\mathrm{Sr}}$) pair to ease the discussion of the Ti$_{i}$+2$\times V_{\mathrm{Sr}}$ case, which will then be Ti$_{\mathrm{Sr}}$ antisite plus $V_{\mathrm{Sr}}$ defect complex (Ti$_{\mathrm{Sr}}$+$V_{\mathrm{Sr}}$).   

Previous theoretical works based on density functional theory (DFT) predicted the Ti$_{\mathrm{Sr}}$ antisite alone to induce a switchable spontaneous polarization in STO~\cite{Choi+09prl,Lee+15Sc,Klyukin+17prb}. Based on these reports, the Ti$_{\mathrm{Sr}}$ antisite is suggested as the origin of the experimentally observed $PS$ and $RS$~\cite{Lee+15Sc,Yang+15apl,Baki+21sr}, although its exact role remains unclear. Specifically, this defect acts as an electron donor~\cite{Choi+09prl}, whereas the experimental grown samples~\cite{Baki+21sr} exhibit an insulating character. This suggests that the Ti$_{\mathrm{Sr}}$ antisite is compensated by $V_{\mathrm{Sr}}$. Therefore, detailed knowledge of the defect characteristics in such a prototype material, including the unexplored defect complex (Ti$_{\mathrm{Sr}}$ + $V_{\mathrm{Sr}}$, see below), is crucial toward tailoring their role and developing promising applications based on the $PS$ and $RS$ behaviors.
 
So far, from the computational side, mostly semi-local and/or Hubbard functionals were used to explore the Ti$_{\mathrm{Sr}}$ antisite~\cite{Choi+09prl,Liu+14pccp,Lee+15Sc,Klyukin+17prb}. A few studies have employed the hybrid exchange-correlation (XC) functional HSE06 for both energy and geometry relaxation of the defects, but either with small supercells (160-atoms)~\cite{Daesu+pem18} or with supercells that exclude/constrain the octahedral tilt of the proveskite lattice (see Fig.1 below)~\cite{Janotti+14prb}. A previous report demonstrated the complexity and the importance of considering a large supercell also accommodating the octahedral tilt, spin-polarization, and a careful consideration of the XC effects to properly describe the characteristics of intrinsic defects~\cite{Marzari+20prs}. Defect complexes, which we intend to explore here, could uncover more challenges, and therefore necessitate thorough analysis.

In this paper, motivated by the experimentally observed $PS$ and $RS$ behaviors~\cite{Lee+15Sc,Yang+15apl,Mikheev+15sr,Baki+21sr} as well as the challenges from the theoretical predictions, we aim to carefully explore the characteristics of defects in STO, including the unexplored defect complex ''Ti$_{\mathrm{Sr}}$ antisite + $V_{\mathrm{Sr}}$''. With the recent remarkable progress in the performance of the FHI-aims~\cite{FHI-aims} package achieved with hybrid functionals~\cite{Kokott+24jcp}, we employ DFT with the Heyd-Scuseria-Enzerhof (HSE06)~\cite{HSE06} XC-functional for all calculations, including the structural relaxation, unless otherwise stated. At first we will discuss the different STO phases/symmetries and report the stability diagram as function of the chemical potential of the components. Then the thermodynamic stabilities of different intrinsic point defects such as vacancies, interstitials, and their complexes, are analyzed. We demonstrate the critical role of the XC treatment in predicting the defect characteristics. We find that the Ti$_{\mathrm{Sr}}$ antisite is energetically favorable and acts as an electron donor. Remarkably, an additional $V_{\mathrm{Sr}}$, which acts as an electron acceptor, enhances the Ti$_{\mathrm{Sr}}$ stability and compensates its charge. This defect complex is found to exhibit a spontaneous $P$ comparable to typical ferroelectrics such as BaTiO$_3$. By switching the defect position between equivalent sites, we explore possibilities and energy barriers for switching the polarization along different pathways. We also discuss how the switching affects the electronic properties of the material. 

\section{METHOD}

All calculations are performed using FHI-aims~\cite{FHI-aims}, an all-electron full-potential package employing numerical atom-centered orbitals. For XC effects, the hybrid functional HSE06~\cite{HSE06} is considered for all calculations (structural relaxations and electronic properties). For all atomic species, we use an \textit{intermediate} setting. The generalized gradient approximation (GGA) in the PBEsol parameterization \cite{PBEsol} is considered, in selected cases, to demonstrate the important role of the XC functional. In the latter, a \textit{tight} setting is used, but \textit{intermediate} already converges both band gap and lattice parameter up to 20~meV and 0.003~\AA, respectively. 

For pristine STO, cubic with a 5-atom unitcell and tetragonal with a 10-atom unitcell are considered. Tetragonal with a 20-atom $\sqrt{2}\times\sqrt{2}\times2$  structure is also considered for exploring other symmetries. The sampling of the Brillouin zone (BZ) is performed with a $\sim$0.2~\AA$^{-1}$ grid density for these pristine unit cells (\textit{i.e.} an 8$\times$8$\times$8 $\textbf{k}$-grid for the 5-atom cubic structure). Lattice constants and internal coordinates are optimized until the residual forces on each atom are less than 0.001~eV/\AA. For these bulk phases, HSE06 incorporates $\alpha$=25$\%$ of Hartree-Fock exact-exchange [HSE(25) hereafter].

For point defects simulation, we use a 320-atom 4$\times$4$\times$4 supercell, which can accommodate the octahedral tilts. At this size, the interactions between the periodic images of different cells are negligible, as reported in Ref.~\onlinecite{Marzari+20prs}. The HSE06 functional is also used for atomic relaxation of all supercells, but with $\alpha$=20$\%$ [HSE(20) hereafter], which is tuned to set the band gap within the range of the experimental values (see below). The reason for using $\alpha$=25$\%$ in pristine STO unitcell is also discussed. BZ integration is performed on a $\Gamma$-centered 1$\times$1$\times$1 $\textbf{k}$-grid. This choice converges the defects' transition-energy level up to 50 meV compared with a $\Gamma$-centered 2$\times$2$\times$2 $\textbf{k}$-grid, as tested at the PBEsol level (with 320-atom cell) for the defect complex. Lattice parameters are fixed to those of a defect-free supercell, while all atomic positions in the supercell are optimized ($\textless$0.005~eV \AA$^{-1}$). Spin polarization is considered in both relaxation and electronic properties for all defects calculations. 

For simulating collective movement with high defects concentration (12.5$\%$), a 40-atom 2$\times$2$\times$2 supercell is considered along with a 4$\times$4$\times$4 $\textbf{k}$-grid sampling. The defects are homogeneously distributed. In this case, both internal coordinates and lattice parameters are optimized ($\textless$0.001~eV \AA$^{-1}$). Their band structures are computed with an 8$\times$8$\times$8 $\textbf{k}$-grid. To evaluate the energy barriers for switching the defect positions, the Nudged Elastic Band method~\cite{Henkelman+20jcp} can be used. However, as the lattice parameters change along the switching paths, we instead create intermediate \textit{images} with linear interpolation of the atomic coordinates. Then, for each image, we only fix the Ti$_{\mathrm{Sr}}$ position and the Sr atom at the origin (\textit{i.e.} fix one Sr<->Ti$_{\mathrm{Sr}}$ distance) and allow all other atomic positions and lattice parameters to relax. This allows to approach the lowest energy path. The changes in the lattice parameters upon switching is discussed below. The \textit{x}-axis in the energy-barrier plots is the distance (\textit{d}) between the \textit{initial} and \textit{final} Ti$_{\mathrm{Sr}}$ position.

The magnitude of the polarization induced in a given structure is evaluated within the Berry phase approach~\cite{berryPhase+93prbr} and referenced to its perfect structure, as shown in the supporting information (SI). This requires the calculation of the polarization branch going from the centrosymmetric cubic to the tetragonal to the fully relaxed defect structure. The intermediate \textit{images} are generated with linear interpolation of the atomic coordinates. The \textit{x}-axis in the polarization plots shown in the SI represents the mass-weighted configurational coordinate (Q) of each \textit{image}, referenced to the perfect structure (Q=0). This is evaluated using the \textit{nonrad} package~\cite{nonrad}.
 
Vacancies (interstitials) are created by removing (adding) the corresponding atomic species. Charged defects states are obtained by adjusting the number of electrons. In these cases, a homogeneous background charge density is introduced for compensation. The formation energy ($E_{f}$) of a defect is given by:
\begin{equation}
E_{f}^{D^{q}}= E_{tot}^{D^{q}}-E_{tot}^{\mathrm{bulk}}-\sum_{i}n_{i}\mu_{i}+q(E_{\mathrm{F}}+E_{\mathrm{VBM}}^{\mathrm{bulk}})+ E_{\mathrm{corr}}^{q}
\end{equation}
where $E_{tot}^{D^{q}}$ is the total energy of the supercell containing a defect $D$ in charge state $q$; $E_{tot}^{\mathrm{bulk}}$ that of the defect-free counterpart; $n_{i}$ is the number of atoms being removed ($\textless$0) or added ($\textgreater$0) to create the defect $D$, and $\mu_{i}$ is the corresponding chemical potential. The latter,  is defined as $\mu_{i}=\Delta\mu_{i}+\mu^{\mathrm{ref}}_{i}$. Thus, it is referenced to the total energy per atom of a physical reservoir such as a gas or a bulk phase. The electron chemical potential (Fermi-level, $E_{\mathrm{F}}$) can be in the range between the valence band maximum (VBM) and the conduction band minimum (CBM) of the perfect STO supercell. The term $E_{\mathrm{corr}}^{q}$ accounts for the finite-size effects as well as for the correct alignment of the electrostatic potential between the defected and the pristine system. Negligible values are found from the latter correction, while the former is ignored because of the high dielectric constant ($\sim 10^{4}$~\cite{Burkard+79prb}) and the large supercell considered. Thus, $E_{\mathrm{corr}}$ is neglected.
 
The formation energy of bulk SrTiO$_{3}$ is defined in terms of the atomic chemical potentials as $ \Delta H_{\mathrm{f}}^{\mathrm{SrTiO_{3}}}=\Delta\mu_{\mathrm{Sr}}+\Delta\mu_{\mathrm{Ti}}+3\Delta\mu_{\mathrm{O}}$. The chemical potentials of the different elements are defined as $$ \mu_{\mathrm{Sr}}=\Delta\mu_{\mathrm{Sr}}+E_{\mathrm{Sr}}^{\mathrm{bulk}},\mu_{\mathrm{Ti}}=\Delta\mu_{\mathrm{Ti}}+E_{\mathrm{Ti}}^{\mathrm{bulk}},\mu_{\mathrm{O}}=\Delta\mu_{\mathrm{O}}+\frac{E_{\mathrm{O_{2}}}}{2},$$
where $E_{\mathrm{Sr}}^{\mathrm{bulk}}$, $E_{\mathrm{Ti}}^{\mathrm{bulk}}$, and $E_{\mathrm{O_{2}}}$ are the total energies of the Sr, Ti, and O atoms in their bulk reference stable states (FCC, hexagonal-$P6_{3}/mmc$, and molecular phase, respectively). To plot the STO energy diagram as a function of $\Delta\mu_{\mathrm{Ti}}$ and $\Delta\mu_{\mathrm{O}}$, we consider the boundary conditions related to the formation of SrO and TiO$_{2}$ precipitations, as follows: $\Delta\mu_{\mathrm{Sr}}+\Delta\mu_{\mathrm{O}}<\Delta H_{\mathrm{f}}^{\mathrm{SrO}}$ and $\Delta\mu_{\mathrm{Ti}}+2\Delta\mu_{\mathrm{O}}<\Delta H_{\mathrm{f}}^{TiO_{2}},$ where $\Delta H_{\mathrm{f}}$ is the formation enthalpy of each precipitation. The calculated $\Delta H_{\mathrm{f}}$ of SrTiO$_{3}$, SrO (rocksalt), and TiO$_{2}$ (anatase, rutile also have a similar value) are about -16.66, -5.69, and -9.60~eV/formula-unit (f.u.), respectively.  
The boundaries given above are expressed as a function of $\Delta\mu_{\mathrm{Ti}}$ and $\Delta\mu_{\mathrm{O}}$, as follows
\begin{equation} 
\left\{ \begin{array}{l}\Delta\mu_{\mathrm{Ti}}+2\Delta\mu_{\mathrm{O}}>\Delta H_{\mathrm{f}}^{\mathrm{SrTiO_{3}}}-\Delta H_{\mathrm{f}}^{\mathrm{SrO}},\\
\Delta\mu_{\mathrm{Ti}}+2\Delta\mu_{\mathrm{O}}<\Delta H_{\mathrm{f}}^{\mathrm{TiO_{2}}}

\end{array} \right.
\end{equation}  

Atomic structures are visualized using the VESTA software \cite{VESTA}. 
\section{RESULTS AND DISCUSSION}
%
\begin{figure}[b!]
 \begin{center}
\includegraphics[width=.49\textwidth]{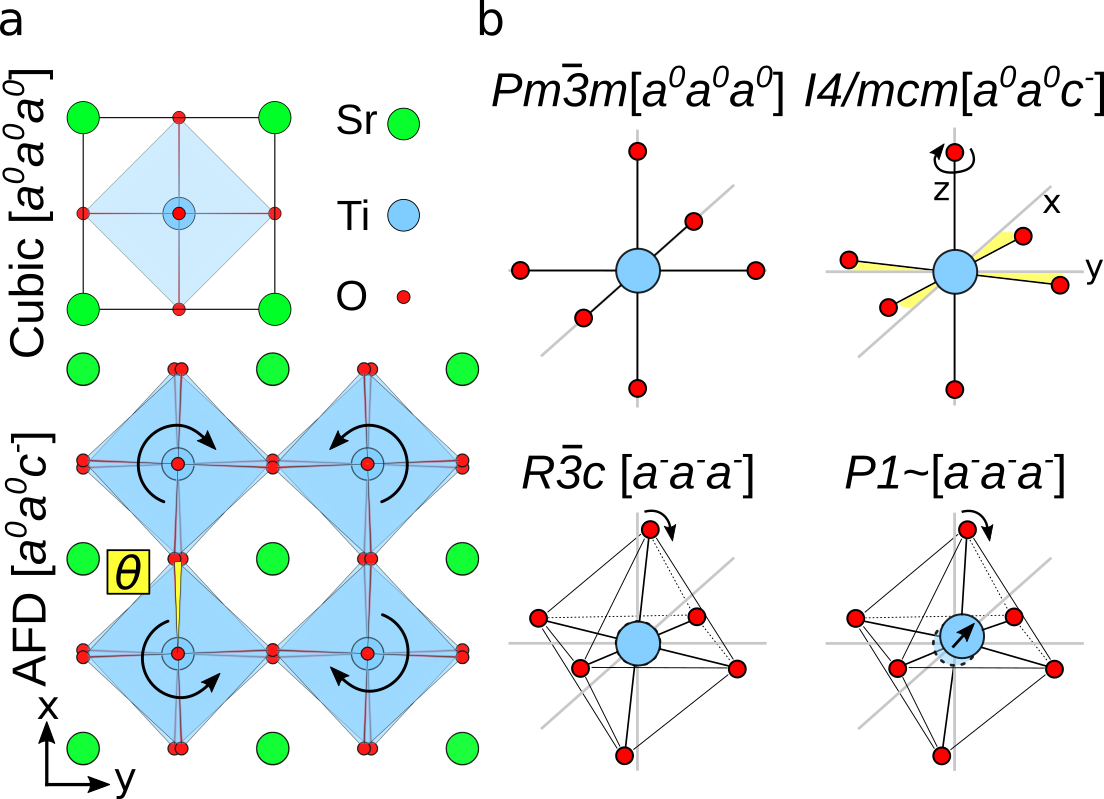}%
\caption{(a) Top view of the cubic and the anti-ferrodistortive (AFD) structures reported experimentally at room and low temperature, respectively. The yellow color highlights the tilt angle $\theta$ of the AFD structure in the top view. (b) Sketch of the TiO$_{6}$ octahedral in the cubic ($Pm\overline{3}m$) and the AFD ($I4/mcm$, $R\overline{3}c$, and $P1$) geometries shown in Table 1. In the $P1$ configuration, the Ti atoms are also shifted along the [111] direction as highlighted by black arrow.}
\label{fig:LaIn}
 \end{center}
\end{figure}
%

\noindent\textit{Characteristics of pristine SrTiO$_{3}$}

As our goal is to simulate defects in large supercells, it is important to explore the pristine STO phases, the role of the XC treatment, the supercells size, and select the optimal ones to achieve our goal.

Let us first summarize the reported characteristics of the bulk STO, which have been intensively studied by both theory and experiment. At room temperature, STO crystallizes in the centrosymmetric cubic phase ($Pm\overline{3}m$ space group), making it a non-ferroelectric material [Fig. 1a]. The experimentally measured lattice parameter is about 3.905~\AA~\cite{Cao+20pssa}. At low temperature ($\textless$110 K) STO exhibits a tetragonal phase~\cite{Unoki+67jpsj,Fleury+68prl}, which is described as an anti-ferrodistortive (AFD) structure with a space group $I4/mcm$ and a tilt mode $a^{0}a^{0}c^{-}$ in the Glazer notion~\cite{Unoki+67jpsj,Glazer+72acsb}. As shown in Fig. 1a, the neighboring TiO$_{6}$ octahedra tilt in an opposite and collective way around the \textit{z} axis, compensating the induced polarization of each others, making STO paraelectric. Experimentally, ferroelectricity in the AFD structure only emerges upon applied strain~\cite{Li+06prb}. The measured structural parameters~\cite{Unoki+67jpsj,Cao+20pssa} are shown in Table 1. Notably, $a$ differs from the $c$ parameter and a tilt angle of 2.1$^{\circ}$ is observed~\cite{Unoki+67jpsj}. Soft phonon modes, which describe the collective movement of atoms, are found as the trigger of such symmetry-lowering distortions~\cite{Fleury+68prl,Cowley+69ssc,Yamada+69jpsj}. Their frequencies vary with temperature~\cite{Fleury+68prl,Cowley+69ssc,Yamada+69jpsj} and, consequently, above $\sim$110 K STO undergoes a phase transition to cubic~\cite{Scott+74rmp}.

Looking now into the electronic properties, the reported experimental optical gap of bulk cubic STO is between 3.10--3.25~eV~\cite{Bieger+92ssi,Galazka+15pss,Benthem+01jap,Schmidt+16prb} at room temperature, 3.22--3.26~eV at $\sim$110K (just above the transition), and 3.2--3.27~eV~\cite{Bieger+92ssi,Galazka+15pss,Schmidt+16prb} for the low temperature phase ($I4/mcm$). These experimental uncertainties can be attributed to both samples' quality and characterization techniques~\cite{Bieger+92ssi,Galazka+15pss,Benthem+01jap,Schmidt+16prb}. As the magnitude of the excitonic effect in STO (20--120 meV~\cite{Schmidt+16prb,Varrassi+21prm}) is within the optical gap uncertainties, in the following we directly comment/compare our electronic band gap to the experimental optical gap. We ignore corrections from the zero-point vibration as we aim to reproduce the experimental gaps, which will be used later for the defects' energy diagram.

\begin{table*}[t]
\centering
\caption{Structural parameters and electronic properties of the cubic and the AFD phases of STO. Space group (s.g.), lattice parameter ($a$), $c/a$ ratio, tilt angle ($\theta$), the difference with the cubic phase energy ($\Delta E$), polarization $P$, and electronic band gap are shown for all symmetries using different XC functionals. The experimental values are shown for the corresponding tilt modes only.}
\vspace{0.2cm}
 \begin{tabular}{c|c|c|c|c|c|c|c|c|c}
\hline
\hline
Phase & Tilt & s.g. & XC & $a$ & $c/a$ & $\theta^{\circ}$ & $\Delta E$ meV/$f.u.$ & $P\mu C/cm^{2}$ & Band gap\\
\hline
\multirow{5}*{Cubic} & \multirow{5}*{$a^{0}a^{0}a^{0}$} & \multirow{5}*{Pm$\overline{3}$m}&PBEsol & 3.889 &1&0&0&0&1.83 \\
 \cline{4-10}
           & &  & HSE06(20$\%$) & 3.904 &1&0&0&0&2.95 \\
 \cline{4-10}
           & &  & HSE06(25$\%$) & 3.896 &1&0&0&0&3.28  \\
\cline{4-10}         
           & &  & Exp.($\sim$110 K) & \, 3.900$^a$ &1&0&-&0& 3.22--3.26$^b$ \\ \cline{4-10}         
           & &  & Exp.(300 K) & \, 3.905$^a$ &1&0&-&0& 3.10--3.25$^b$ \\
 \hline
\multirow{10}*{AFD}&\multirow{4}*{$a^{0}a^{0}c^{-}$} & \multirow{4}*{$I4/mcm$}&PBEsol & 3.879 & 1.0062 & 5.6 & -7.7 & 0 & 1.94 \\
\cline{4-10}
  &  & & HSE06(20$\%$) & 3.901 & 1.0018 & 2.9 &-1.1&0&2.98 \\
\cline{4-10}
  &  & & HSE06(25$\%$) & 3.895 & 1.0010 & 2.2 &-0.8&0&3.29 \\
\cline{4-10}
  &  & & Exp.($\textless$66 K) & \, 3.898$^a$ & \, 1.0006$^a$ & \, 2.1$^c$ &-&-&3.2--3.27$^b$ \\
\cline{2-10}
&\multirow{3}*{$a^{-}a^{-}a^{-}$} & \multirow{3}*{$R\overline{3}c$} & PBEsol & 3.887& 1 & 4.3 & -7.6 & 0 & 1.96 \\
\cline{4-10}
& & &HSE06(20$\%$)& 3.904 & 1 & 2.1 & -2.3 & 0 & 2.99\\
\cline{4-10}
& & &HSE06(25$\%$)& 3.896 & 1 & 1.5 & -2.1 & 0 & 3.29\\
\cline{2-10}
&\multirow{3}*{$\sim a^{-}a^{-}a^{-}$} & \multirow{3}*{$P1$} & PBEsol & \! $\sim$3.891  & 1 & $\sim$4.5 & -9.6 & 22.7  & 2.11\\
\cline{4-10}
 & & & HSE06(20$\%$) & 3.911 & 1 & 3.8 & -6.2 & 21.3 & 3.26 \\
\cline{4-10}
& & & HSE06(25$\%$) & 3.902& 1 & 3.2& -4.3 & 22.1 & 3.52 \\
\hline
\hline
\end{tabular}
\begin{flushleft}
$^a$ Reference~\cite{Cao+20pssa} at 66, 115 and 300 K.\\
$^b$ Experimental optical gaps~\cite{Bieger+92ssi,Galazka+15pss,Benthem+01jap,Schmidt+16prb} at ~0K, ~110K, and 300K.\\
$^c$ Reference~\cite{Unoki+67jpsj} at 4.2 K.
\end{flushleft}
\end{table*}

From previous theoretical reports, AFD phase ($I4/mcm$) with collective $a^{0}a^{0}c^{-}$ tilt mode is predicted to exhibit lower energy (few meV/f.u) than the cubic structure~\cite{Wahl+08prb,Mellouhi+11prb}. Such an energy gain is mainly from the octahedra tilt. Other tilt modes such as $a^{-}a^{-}a^{-}$ and $a^{0}b^{-}b^{-}$ were also predicted theoretically to lower the energy and stabilize the tetragonal phase~\cite{Rosendal+23prm}. While STO shows a competition between the AFD and the Ferroelectric instability, cooperation between them can occurs at larger tilt angles, leading to a ferroelectric state~\cite{Spaldin+14jpcm}. Previous theoretical work, using hybrid functional HSE06, predicted ferroelectric AFD structure~\cite{Watanabe+19cms} when considering a large cell (20-atom supercell). It is mainly attributed to the off Wyckoff displacement of the Ti-lattice. Therefore, to properly simulate the defect-induced properties (\textit{e.g} polarization), it is important to carefully check the characteristics of larger STO supercells without constraining their symmetry.

Now we proceed with exploring the characteristics of both STO phases and selected tilt modes, based on the above reports. We also consider a large pristine-AFD supercell. In Table 1, we report the structural parameters, the energy again of the AFD phase, the polarization, and the band gap for the cubic ($Pm\overline{3}m$) as well as the AFD ($I4/mcm$, $R\overline{3}c$, and $P1$) phases [see Fig. 1b]. The experimental counterparts are shown for comparison. We consider both PBEsol and HSE06 functionals. As we do not know how the symmetry-lowering in the AFD phase affects the band gap, we consider HSE(20) in addition to the standard HSE(25).  

Looking first into the cubic phase, the calculated lattice constant of 3.896~\AA, obtained with the HSE(25) functional, is in good agreement with the experiment~\cite{Cao+20pssa}. The PBEsol counterpart is slightly lower, being 3.889~\AA (see Table 1). Similar behavior is found for the AFD($I4/mcm$) structure with the $a^{0}a^{0}c^{-}$ tilt. The HSE(25) functional produces the lattice parameter, the $c/a$ ratio and, importantly, the tilt angle (2.2$^{\circ}$) very well. The latter is largely overestimated by PBEsol (5.6$^{\circ}$). This AFD configuration is lower in energy compared to the cubic phase by about 7.7 meV/$f.u$ as obtained with PBEsol. As the lattice parameter of the AFD and the cubic are close, the energy gain is due to the octahedra tilt. The latter value deceases to about 1 meV/$f.u.$ with HSE(25), reflecting the well-predicted tilt angle.

Our calculations tell that the $a^{-}a^{-}a^{-}$ tilt mode ($R\overline{3}c$) is lower in energy compared to the $a^{0}a^{0}c^{-}$ structure [see Table 1]. Considering 20-atom $\sqrt{2}\times\sqrt{2}\times2$ supercell lowers the symmetry of $a^{-}a^{-}a^{-}$ tilt modes ($P1$) as it allows more atomic distortions, including off-center shifts [see Fig. 1b]. This configuration is the lowest in energy and exhibits a spontaneous polarization of 22 $\mu C/cm^{2}$, mainly from the off-center shift of the whole Ti-lattice in the [111] direction. As such, the TiO$_{6}$ octahedral exhibits three shorter (1.89\AA) and three elongated (2.02\AA) Ti-O bonds, unlike the cubic, $I4/mcm$ and $R\overline{3}c$ where all Ti-O bond length are about 1.95 \AA~[Fig. 1b]. Similar characteristics are found with PBEsol, but with a larger $\Delta E$ compared to HSE(25), because of the overestimated~$\theta$. The polarization in the PBEsol and the HSE(25) structures is about 22.7 and 22.1 $\mu C/cm^{2}$, respectively. We note that for a given structure, the use of PBEsol or HSE06 functional to compute the polarization gives comparable results, as the VB states are almost similar with both functionals in the pristine STO case. However, the small differences in the polarization shown in Table 1 come from the fact that the HSE(25) and PBEsol geometries structurally differ. Up to now, we see that considering larger supercell, which is needed later to simulate point defects, affects the crystal symmetry and induces spontaneous polarization.  

Let us briefly discuss the effect of including the van der Waals (vdW) correction on the AFD($P1$) geometry. To this end, we considered the Tkatchenko-Scheffler~\cite{vdWTS+09prl} (TS) and the many-body dispersion~\cite{MBD+12prl} corrections. We found that both reduce the atomic distortions (and thus the polarization) of the $P1$ structure, making it close to $R\overline{3}c$ symmetry. However, the lattice parameter is largely underestimated, being 3.84 and 3.86 \AA, respectively, indicating that both give too strong vdW correction. Underestimated lattice parameters largely affect the energy and electronic properties of STO. Therefore, the HSE06 functional alone is considered the optimal choice for our study. 

Comparing the electronic band gaps, we clearly see that HSE(25) is more accurate at predicting the experimental gaps of bulk STO phases ($Pm\overline{3}m$, $I4/mcm$, and $R\overline{3}c$) compared to PBEsol. We found that HSE(20) gives comparable structural parameters, but largely underestimates the band gap, being 2.95 (cubic) and 2.98/2.99~eV ($I4/mcm$/$R\overline{3}c$). However, for the AFD $P1$ structure, the HSE(25) band gap is, instead, largely overestimated (3.52 eV). This increase compared to $Pm\overline{3}m$, $I4/mcm$, and $R\overline{3}c$ structures is mainly due to the symmetry-lowering distortions discussed above which emerge in the $P1$ geometry. In this peculiar case, the HSE(20) is more accurate at predicting the experimental gap, being 3.26 eV. The structural parameters obtained from the HSE(20) are in close agreement with both the HSE(25) and the experiment [see Table 1].  This is an example where tuning the  $\alpha$ parameter gives minor changes in the geometry but largely affects the band gap. 
 
Overall, based on the above results, a 4$\times$4$\times$4 supercell with AFD $P1$ ($a^{-}a^{-}a^{-}$) phase, which accommodates the octahedral tilt and other distortions, will be considered for exploring the point defects. Simulating defects with this 0 K phase will not affect our goal to understand the defect characteristics in STO. Specifically, the $PS$ and $RS$ phenomena are also observed at low temperatures (\textit{e.g.} 10K~\cite{Baki+21sr}). Also, the use of the lowest-energy $P1$ structure instead of $I4/mcm$ and $R\overline{3}c$ is because one cannot constrain the symmetry upon introducing defects. Evaluating the polarization induced by defects is also unaffected as we will subtract (refer to) the polarization of the pristine $P1$ (see SI). Similar to the $P1$ ($a^{-}a^{-}a^{-}$) structure shown in Table 1, HSE(20) will be considered for all calculations of defects. This gives a band gap of $\sim$3.3~eV for the 4$\times$4$\times$4 defect-free AFD supercell, which will be considered as a reference for the defect formation energy diagram. 


\begin{figure*}[t!]
 \begin{center}
\includegraphics[width=.99\textwidth]{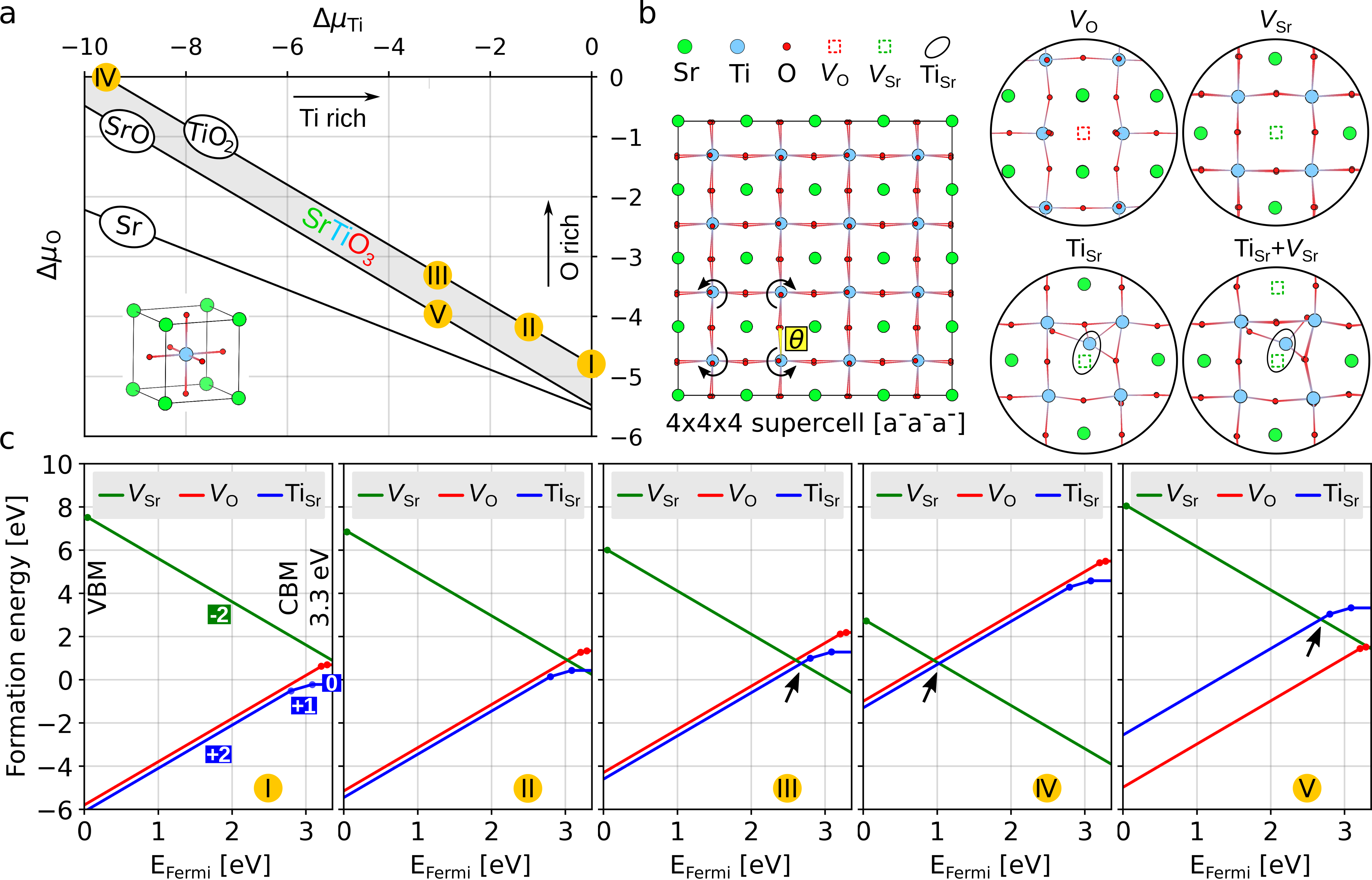}%
\caption{(a) Stability energy diagram of bulk STO in the cubic phase as a function of the chemical potential of titanium and oxygen, in~eV. The STO phase can be formed within the shaded grey area, comprised between the boundaries (solid lines) of TiO$_{2}$ and SrO precipitations. (b) Sketch of the AFD supercell and the defects considered in this work. (c) Defects' formation energy as a function of the Fermi level. The slopes of the lines reflect the charge state $q$. The dots represent the thermodynamic transition level between them [for explicit numbers see panel (c) left]. Each panel refers to a corresponding chemical potential condition highlighted in (a) by numbered orange circles.}
\label{fig:LaIn}
 \end{center}
\end{figure*}
%

\noindent \textit{Point defects characteristics} 

In Fig. 2, we depict a comprehensive compilation of the stability of selected intrinsic point defects in STO. The selection is based on previous theoretical and experimental reports of $PS$ and $RS$~\cite{Choi+09prl,Klyukin+17prb,Lee+15Sc,Yang+15apl,Mikheev+15sr,Baki+21sr}. We consider $V_{\mathrm{O}}$, $V_{\mathrm{Sr}}$, Ti$_{\mathrm{Sr}}$ antisite as well as defect complex Ti$_{\mathrm{Sr}}$+$V_{\mathrm{Sr}}$. These defects are introduced within a 4$\times$4$\times$4 supercell with AFD a$^{-}$a$^{-}$a$^{-}$ tilt mode as mentioned above [see Fig. 2b]. 
 
Looking first at the stability energy diagram, the STO cubic phase can be formed within the area limited by the SrO and TiO$_{2}$ boundaries [see Fig. 2a]. A similar diagram is found when using PBEsol instead of HSE06 and/or AFD instead of the cubic phase. The experimental reports of the $RS$ in STO tuned the stoichiometry of the samples going toward Ti-\textit{rich} and Sr-\textit{poor}~\cite{Baki+21sr}. Considering this condition, we selected four points ($I$, $II$, $III$, $IV$) to compute the corresponding defect stability, which reflect the \textit{rich}-to-\textit{poor} limit of Ti, but, within the TiO$_{2}$-\textit{rich} conditions [see Fig. 2a]. An additional point, with SrO-\textit{rich} ($V$), is also considered for comparison/prediction. 

Focusing now on the defect stability, at point ($I$) the Ti$_{\mathrm{Sr}}$ antisite exhibits the lowest formation energy [Fig. 2c]. This is expected to be abundant, inline with experimental observations~\cite{Karjalainen+20jap,Baki+21sr}. The Ti$_{\mathrm{Sr}}$ antisite is stable in a +2 charge state (Ti$_{\mathrm{Sr}}^{+2}$) [Fig. 2c]. Increasing the Fermi level, a +2/+1 thermodynamic transition appears at about 2.8~eV, which is 0.50~eV below the electronic band gap of the pristine supercell ($\sim$3.3~eV). Here, the defect captures one electron. The Ti$_{\mathrm{Sr}}^{+1}$ configuration exhibits a magnetic moment of 1$\mu_{B}$. This defect state is suggested as the origin of the broad luminescence peak reported experimentally at 1.8-1.9~eV~\cite{Daesu+pem18,Noesges+22jvst}, considering the reported strong Stokes shift of $\sim$0.9~eV~\cite{Daesu+pem18}. Our results indicate that the Ti$_{\mathrm{Sr}}$ defect is a deep center, but closer to CBM than reported previously (transitions at 2.1~eV~\cite{Choi+09prl} and 2.6~eV~\cite{Daesu+pem18}). These differences are attributed to the use of Hubbard correction functional along with \textit{odd}-size supercell in the former, which excludes the octahedra tils~\cite{Choi+09prl}, and the use of a smaller supercell (160-atoms) in the latter~\cite{Daesu+pem18}. Although hybrid functional is used in the latter, the supercell size is reported to affect the transition point~\cite{Marzari+20prs}. However, our conclusions agree in terms of the transition type and behavior~\cite{Choi+09prl,Daesu+pem18}, compared with reports done at the semi-local functional level where this transition point is predicted, instead, close to VBM~\cite{Liu+14pccp}. This is mainly due to the fact that the captured electron is hosted by Ti-\textit{d} orbital. At 3.08~eV where a +1/0 transition occurs, the Ti$_{\mathrm{Sr}}$ antisite captures two electrons, becoming neutral. This transition (0.22~eV below the CBM) is higher than the room thermal energy $k_{B}T$, indicating that Ti$_{\mathrm{Sr}}$ is still a deep donor. The neutral Ti$_{\mathrm{Sr}}$ configuration exhibits a magnetic moment of 2$\mu_{B}$, where both electrons are mainly captured by Ti-\textit{d} orbital, with negligible contribution from oxygen.   

Analyzing the antisite geometry, we clearly see that the Ti atom shifts from the $V_{\mathrm{Sr}}$ position (off-center) along the [$\sim$110] direction [Fig. 2b]. The shift is accompanied by structural distortions, reflecting its deep donor character. This includes a shift of the neighboring Ti atoms, distortions of the O atoms (thus the octahedral tilts), and formation of bounds with neighboring O atoms. 

In the case of $V_{\mathrm{Sr}}$, we see that it exhibits a high formation energy compared to Ti$_{\mathrm{Sr}}$. The transition points 0/$-$1 and $-$1/$-$2 occur at about 0.03 and 0.12~eV above the VBM, respectively. Its formation energy decreases with the Fermi level ($V_{\mathrm{Sr}}^{-2}$). This makes it a double acceptor with shallow character (transition point near the VBM). This character is reflected in its structure, which shows minor distortions [Fig. 2b]. Such a defect can act as a compensation center in n-type doped STO or with other intrinsic donor defects. Considering the $V_{\mathrm{O}}$, the +2/+1 and +1/0 transition points are found at 3.21 and 3.28~eV, respectively. Despite the diversity in the previous theoretical and experimental reports, our values are in excellent agreement with the recent careful study of the $V_{\mathrm{O}}$ characteristics~\cite{Marzari+20prs}. In the latter report, the role of the computational parameters and functional is found to be the origin of the broad results. Here, considering all key parameters such as spin, supercell size, and accommodating octahedral tilt along with relaxation with HSE(20), we found that the $V_{\mathrm{O}}$ is still a shallow donor inline with this report~\cite{Marzari+20prs}. The $V_{\mathrm{O}}^{+1}$ configuration exhibits a magnetic moment of 1$\mu_{B}$, while the neutral $V_{\mathrm{O}}$ counterpart is non-magnetic. Here, as we refer to the band gap of 3.3 instead of 3.25~eV as explained above, the +1/0 transition point is also a shallow donor, being 0.02~eV below the CBM. Overall, under condition ($I$) Ti$_{\mathrm{Sr}}$ is slightly lower in energy than $V_{\mathrm{O}}$ [Fig. 2b].

Given the opposite character of Ti$_{\mathrm{Sr}}$ antisite and $V_{\mathrm{Sr}}$ (deep donor and shallow acceptor, respectively), as well as the possibility to form under similar conditions (Ti-\textit{rich}, Sr-\textit{poor}), a charge compensation can be achieved. Moving away from the Ti-\textit{rich} limit ($I$ to $IV$), but within the TiO$_{2}$-\textit{rich} line, compensation between the Ti$_{\mathrm{Sr}}^{+2}$ and $V_{\mathrm{Sr}}^{-2}$ is evident starting from point ($III$) [Fig. 2]. From now on, we refer to this Ti$_{\mathrm{Sr}}^{+2}$+$V_{\mathrm{Sr}}^{-2}$ defect complex as compensated Ti$_{\mathrm{Sr}}$+$V_{\mathrm{Sr}}$. Intriguingly, such a complex can be used to pin the Fermi level within the band gap, leading to a semi-insulating material. This conclusion is inline with the insulating character observed in STO films grown under Sr-\textit{poor} (\textit{i.e.} $V_{\mathrm{Sr}}$) and Ti-\textit{rich} (\textit{i.e.} Ti$_{\mathrm{Sr}}$) conditions~\cite{Baki+21sr}. Our result indicates that one can also tune the Fermi level by tuning the chemical potential within the TiO$_{2}$-\textit{rich} boundary.

To validate this analysis and conclusion, we also computed the binding energy of this compensated (Ti$_{\mathrm{Sr}}$+$V_{\mathrm{Sr}}$) defect depicted in Fig. 2b. The binding energy, obtained as the difference between the formation energy of the defect complex minus the formation energy of the individual components (Ti$_{\mathrm{Sr}}^{+2}$ and $V_{\mathrm{Sr}}^{-2}$), is \textit{exothermic}, being about -0.15~eV. We also set the $V_{\mathrm{Sr}}^{-2}$ away from the Ti$_{\mathrm{Sr}}^{+2}$, as a second neighbor, and found an increase of 0.12~eV in the total energy, being less favorable compared to the first-neighbor configuration [Fig. 2b]. Therefore, this compensated defect complex is expected to be abundant in samples grown under Ti-\textit{rich} and Sr-\textit{poor} conditions~\cite{Baki+21sr}.

At point ($II$), this defect complex can, instead, be formed between the neutral Ti$_{\mathrm{Sr}}$ and $V_{\mathrm{Sr}}^{-2}$, giving (Ti$_{\mathrm{Sr}}$+$V_{\mathrm{Sr}}$)$^{-2}$ defect. This charged defect complex can also be achieved at points $III$ and $IV$ by increasing the Fermi level with an external stimulus, \textit{e.g.} under electric voltage. Overall, the compensated defect complex (Ti$_{\mathrm{Sr}}^{+2}$+$V_{\mathrm{Sr}}^{-2}$) has a semi-insulating character, while in the charged counterpart (Ti$_{\mathrm{Sr}}$+$V_{\mathrm{Sr}}$)$^{-2}$ the Ti$_{\mathrm{Sr}}$ component captures two electrons to the Ti-\textit{d} orbital, becoming neutral. 

Moving into the SrO-\textit{rich} condition (point $V$), $\mu_{\mathrm{O}}$ is $poor$ compared to the TiO$_{2}$ limit (point $III$). This makes $V_{\mathrm{O}}$ lower in energy compared to both Ti$_{\mathrm{Sr}}$ and $V_{\mathrm{Sr}}$, which are now shifted to higher energies. This indicates that the compensated defect complex (Ti$_{\mathrm{Sr}}$+$V_{\mathrm{Sr}}$) is unlikely to form in this condition. Also, the presence of $V_{\mathrm{O}}$ can affect its characteristics. Therefore, the TiO$_{2}$-\textit{rich} line [Fig. 2a] is the optimal condition to achieve the defect complex (Ti$_{\mathrm{Sr}}$+$V_{\mathrm{Sr}}$) and benefit/tune its characteristics. This compensating defect complex, formed by the oppositely charged constituents, is expected to show its potential, specifically in inducing a spontaneous polarization. Ti$_{\mathrm{Sr}}$ itself is reported to induce a switchable spontaneous polarization~\cite{Choi+09prl,Lee+15Sc,Klyukin+17prb}. Inline with these reports, the $RS$ observed in STO film was under Sr-\textit{deficiency} and Ti-\textit{excess}. The \textit{deficiency} is estimated to be about 20$\%$ in which 10$\%$ are occupied by Ti atoms, giving Ti$_{\mathrm{Sr}}$ defect, and 10 $\%$ remains $V_{\mathrm{Sr}}$~\cite{Baki+21sr}. Therefore, from now on, we focus on the defect characteristics at higher concentration.
 
\begin{figure}[h!]
 \begin{center}
\includegraphics[width=.49\textwidth]{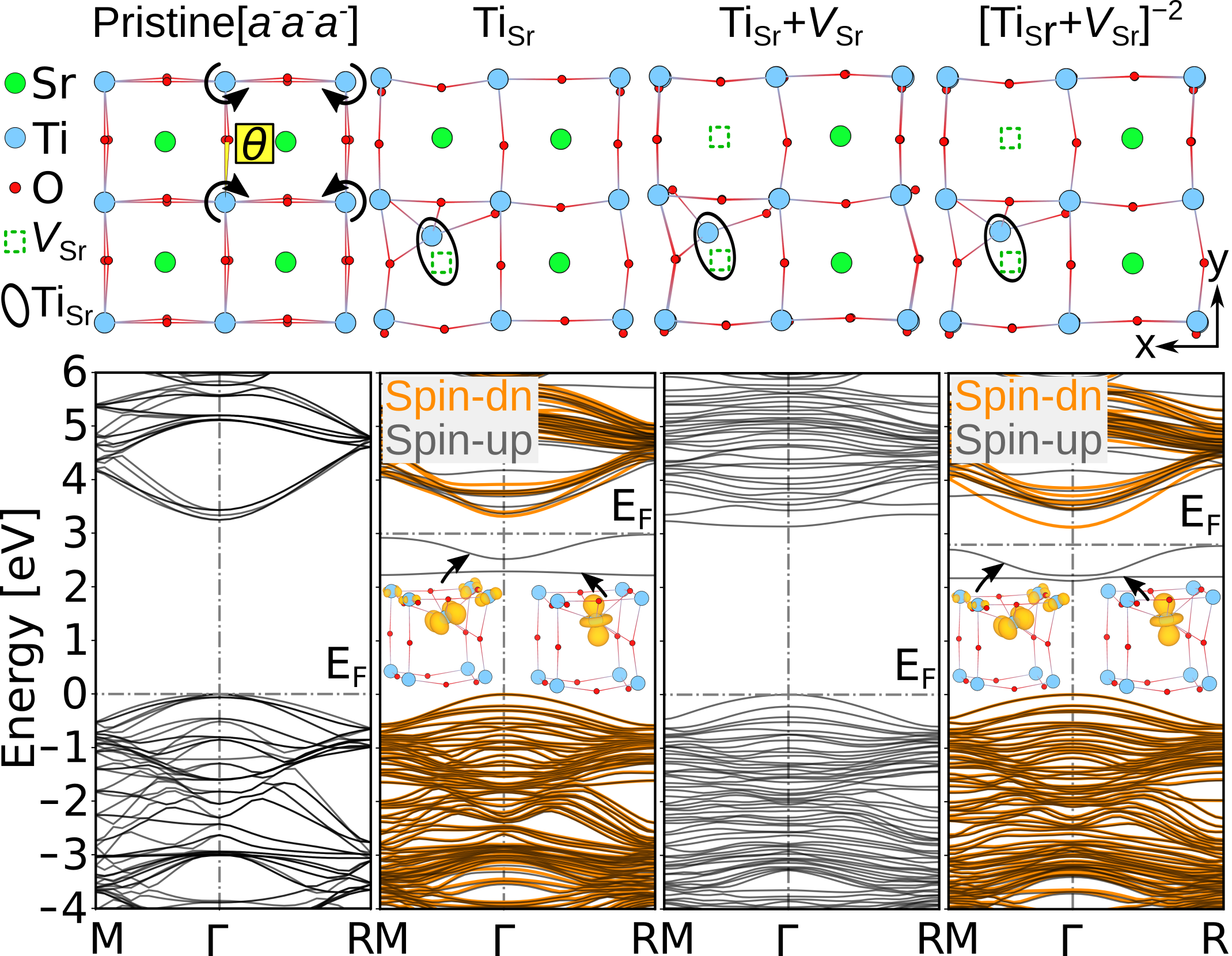}%
\caption{Electronic band structure of (a) pristine STO in $P1$ symmetry (a$^{-}$a$^{-}$a$^{-}$ tilt mode), (b) neutral Ti$_{\mathrm{Sr}}$, (c) compensated Ti$_{\mathrm{Sr}}$+$V_{\mathrm{Sr}}$, and (d) charged (Ti$_{\mathrm{Sr}}$+$V_{\mathrm{Sr}}$)$^{-2}$. The bulk-like valence band maximum is set to zero, while the Fermi level is highlighted by a dashed gray line. The spin up and down are shown with grey and orange colors, respectively. The real space distribution of the defect states is shown within each panel. } 
\label{fig:LaLa}
 \end{center}
\end{figure}

\noindent\textit{Electronic properties of high-defects concentration}

For analysis, comparison, and predictions, in Fig. 3 we present the electronic band structure of the pristine AFD, neutral Ti$_{\mathrm{Sr}}$, compensated (Ti$_{\mathrm{Sr}}$+$V_{\mathrm{Sr}}$) as well as charged (Ti$_{\mathrm{Sr}}$+$V_{\mathrm{Sr}}$)$^{-2}$ defect, each within a 2$\times$2$\times$2 supercell. This corresponds to 12.5$\%$ defect concentration, mimicking the high off-stoichiometry where $PS$ and $RS$ are experimentally observed~\cite{Baki+21sr,Yang+15apl,Lee+15Sc}. Homogeneous defects distribution is considered [Fig. 3]. All configurations are fully relaxed using HSE(20) including the lattice parameters. The structural characteristics of the defects [see Fig. 3] are almost similar to those discussed above for point defects [Fig. 2a]. The Ti$_{\mathrm{Sr}}$ in all cases is shifted within the (110) plane along the [$\sim$110] direction. In the neutral Ti$_{\mathrm{Sr}}$ [Fig. 3], the (110) plane is also the mirror symmetry plane of the structure. In the defect complexes [Fig. 3], the latter symmetry is broken by the distortions of the neighboring Ti atoms. Analyzing the band structures, we see that the neutral Ti$_{\mathrm{Sr}}$ antisite exhibits two mid-gap states, compared to the pristine material [see Fig. 3]. This is due to the fact that the Ti (Sr) atom has 4 (2) valence electrons, thus the Ti$_{\mathrm{Sr}}$ defect brings in two more electrons. Both states exhibit a Ti-$d$ character with spin $up$, and each is occupied by one electron. This gives a total magnetic moment of 2$\mu_{B}$. The lowest- (highest-)energy band exhibits $e_{g}$-like ($t_{2g}$-like) symmetry [Fig. 3b]. While both are very localized in real space, the latter band is slightly dispersed compared to the former, exhibiting contribution from the four neighboring Ti atoms. In pristine STO, the octahedral crystal field splits the Ti-$d$ states in which the $t_{2g}$ is lower in energy. Here, the defect states show the opposite, which can indicate tetrahedral crystal field splitting, making $e_{g}$ lower in energy.

The importance of the XC treatment and the inclusion of spin polarization to properly simulate the neutral Ti$_{\mathrm{Sr}}$ is shown in Fig. S1a in SI. We found that ignoring the spin polarization does not only affect the band structure of Ti$_{\mathrm{Sr}}$ by giving one doubly occupied mid-gap band, but also reduces the geometrical symmetry as the Ti$_{\mathrm{Sr}}$ slightly deviates from the [110] direction (breaking the mirror symmetry plane mentioned above). This geometry is +0.25~eV higher in energy compared to the spin-polarized (2$\mu_{B}$) counterpart [see Fig. S1a in SI]. Remarkably, optimization of the neutral Ti$_{\mathrm{Sr}}$ using PBEsol functional gives a metallic character with two Ti-$d$ electrons at the CBM instead of localized mid-gap states, despite its band structure is computed with HSE(20) [see Fig. S1a]. This example indicates that HSE(20), or at least Hubbard correction, is crucial also for the structural relaxation of defects as it can drastically change the electronic properties. Tuning the Hubbard parameter is challenging for the defect complex considered here (\textit{e.g.} Ti$_{\mathrm{Sr}}$+$V_{\mathrm{Sr}}$) where different states and hybridization are involved. For this reason also, HSE(20) is used for the structural relaxation of all defects. 

For the Ti$_{\mathrm{Sr}}$+$V_{\mathrm{Sr}}$ defect complex, the compensation between the donor Ti$_{\mathrm{Sr}}$ and the acceptor $V_{\mathrm{Sr}}$ preserves the insulating character of the material [Fig. 3c]. The structural relaxations are reflected in the band dispersion compared to its pristine counterpart. We expect this defect complex to be the dominant and the representative type in the reported off-stoichiometric samples where $PS$ and $RS$ are observed~\cite{Baki+21sr,Yang+15apl,Lee+15Sc}. This is not only because of its stability, as discussed above in Fig. 2, but also because the off-stoichiometric films, with $\sim$10$\%$ defect concentration, exhibit an insulating character without applied voltage~\cite{Baki+21sr}. Moving into the charged defect complex (Ti$_{\mathrm{Sr}}$+$V_{\mathrm{Sr}}$)$^{-2}$ [Fig. 3c], this type mimics the situation with increased Fermi level, \textit{e.g.} upon applied voltage, or the case with point $II$ condition, as discussed above [see Fig. 2]. We see that the two electrons are hosted by the Ti$_{\mathrm{Sr}}$ components, showing a similar electronic structure to the individual neutral Ti$_{\mathrm{Sr}}$. Therefore, the $V_{\mathrm{Sr}}$ mainly enhances the stability of the Ti$_{\mathrm{Sr}}$ defects through the charge compensation, and preserves its electronic fingerprint (localized mid-gap states), under \textit{rich} electron chemical potential. 
We also relaxed these defect complexes (both compensated and charged) using PBEsol. We conclude that PBEsol fails only in predicting the geometries of defects that exhibit localized mid-gap states. In the case of a compensated complex (Ti$_{\mathrm{Sr}}$+$V_{\mathrm{Sr}}$), which has a semiconducting character without mid-gap states [Fig. 3], the PBEsol functional successfully predicts its geometry, being similar to that obtained with HSE(20) [see SI]. However, the band structures require HSE(20) in all cases. 
\\

\noindent\textit{Polarization and switching barriers}

As discussed above, the AFD phase ($P1$), which is our pristine reference, exhibits a spontaneous polarization. Thus, to properly determine the polarization induced by the compensated Ti$_{\mathrm{Sr}}$+$V_{\mathrm{Sr}}$ defect within the berry phase approach, we deducted that of the AFD phase. The polarization branches going from the defected centrosymmetric (non ferroelectric) to the AFD into the fully relaxed defect structure are shown in Fig. S2 in SI. These calculated branches are crucial to properly determine the polarization magnitude and direction. We found that the compensated Ti$_{\mathrm{Sr}}$+$V_{\mathrm{Sr}}$ defect induces a polarization of about 32 $\mu C/cm^{2}$ oriented along the Ti-off-center direction [110]. Such a high value is comparable to those of common ferroelectric such as BaTiO$_{3}$ (about 29 $\mu C/cm^{2}$~\cite{Fang+2d}). It reflects the Ti$_{\mathrm{Sr}}$ displacement and the structural distortions around it (displacement of neighboring Ti and O atoms) as well as the strong dipole induced by its components (Ti$_{\mathrm{Sr}}^{+2}$ and $V_{\mathrm{Sr}}^{-2}$).

As depicted in Fig. 4a, from the symmetry point of view we clearly see that the Ti$_{\mathrm{Sr}}$ defect can occupy other equivalent sites, \textit{e.g.} $B$ and $C$. The compensated defect (Ti$_{\mathrm{Sr}}$+$V_{\mathrm{Sr}}$) can also have equivalent sites (\textit{e.g.} $D$) in the neighboring \textit{pseudocube} (formed by the 8 Ti atoms). The energy barriers for switching between the equivalent sites are shown in Fig. 4b and 4c. Looking first at neutral Ti$_{\mathrm{Sr}}$, we found that the switching from $A$ to $B$ faces an energy barrier of about 0.1~eV. While this barrier is larger than the room thermal energy $k_{B}T$, it can be achieved by applying low electric voltage. Following the $A$-$C$ pathway, the barrier increases to about 0.6~eV. From first inspection of the cubic-like geometry, it seems that the $A$-$B$ and $A$-$C$ paths are equivalent. However, considering the oxygen atoms involved in the bonding with the Ti$_{\mathrm{Sr}}$, one can see that the $A$-$C$ pathway requires breaking more bonds compared to the $A$-$B$ path, which can be seen as a sway/oscillation. Also, from the electronic structure view [Fig. 3], the $t_{2g}$-like state is shared with the neighboring four Ti atoms, which are the same along the $A$-$B$ path. However, along $A$-$C$ path, the $t_{2g}$ distribution moves from the four upper Ti atoms to the four bottom Ti atoms. For this reason, the $A$-$C$ swinging pathway requires higher energy (thus higher electric voltage for switching), compared to the $A$-$B$ path. For both paths, we found that all the intermediate \textit{images} exhibit localized mid-gap states with 2$\mu_{B}$, preserving the semiconducting-like character. In the experiment, the films are grown on pristine STO~\cite{Baki+21sr}. This can lead to strain on the films due to its off-stoichiometry. Thus, we also fixed the in-plane lattice parameters of the defected structure to those of pristine STO and found similar behavior in terms of energy barriers, polarization and electronic structure [Fig. 4b].

\begin{figure}[h!]
 \begin{center}
\includegraphics[width=.46\textwidth]{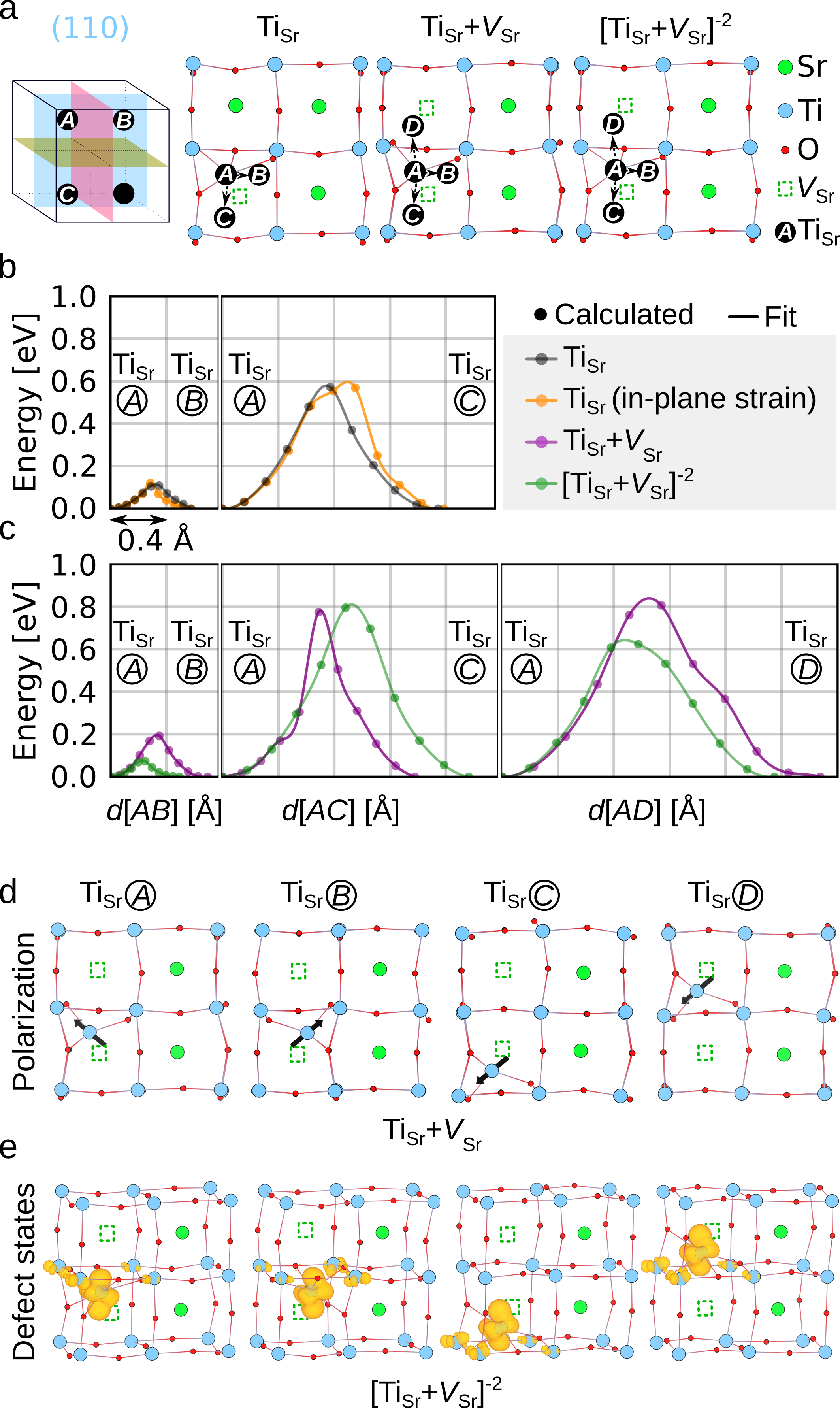}%
\caption{(a) Sketch of the neutral Ti$_{\mathrm{Sr}}$ geometry within a 2$\times$2$\times$2 supercell, corresponding to 12.5$\%$ defect concentration. The (110) plane and the equivalent defect sites considered here are highlighted in the geometry. (b) The energy barriers for switching the position of the neutral Ti$_{\mathrm{Sr}}$ defect along the $A$-$B$ and $A$-$C$ pathways. The \textit{x}-axis in the energy-barrier plots is the distance \textit{d} between the \textit{initial} and \textit{final} Ti$_{\mathrm{Sr}}$ positions along each path. The orange line corresponds to the case where the in-plane lattices are fixed to those of pristine STO. (c) Similar analysis for compensated (Ti$_{\mathrm{Sr}}$+$V_{\mathrm{Sr}}$) and charged (Ti$_{\mathrm{Sr}}$+$V_{\mathrm{Sr}}$)$^{-2}$ defect complexes. Here, an additional path ($A$-$D$) is considered in which the Ti atom moves into the equivalent site in the neighboring \textit{pseudocube}. The latter is defined as the cube formed by the eight neighboring Ti atoms of Ti$_{\mathrm{Sr}}$. The $A$-$D$ path can be seen as a local migration of the Ti$_{\mathrm{Sr}}$. (d) Polarization vector of the compensated defect complex (Ti$_{\mathrm{Sr}}$+$V_{\mathrm{Sr}}$) is highlighted on the Ti atom, upon switching between $A,~B,~C,~D$ sites shown in (a) and (c). (e) For the charged defect complex (Ti$_{\mathrm{Sr}}$+$V_{\mathrm{Sr}}$)$^{-2}$, the sum of the distribution of the occupied mid-gap states induced by the defect [see Fig. 3] is shown for the different sites.}
\label{fig:LaIn-IF}
 \end{center}
\end{figure}

Analyzing the compensated Ti$_{\mathrm{Sr}}$+$V_{\mathrm{Sr}}$ complex, we also see similar characteristics along the $A$-$B$ and $A$-$C$ pathways. The latter has higher barrier ($\sim$0.8~eV) compared to the neutral Ti$_{\mathrm{Sr}}$ ($\sim$0.6), which is attributed to the fact that in the compensated Ti$_{\mathrm{Sr}}$+$V_{\mathrm{Sr}}$ complex the components (Ti$_{\mathrm{Sr}}^{+2}$ and $V_{\mathrm{Sr}}^{-2}$) are oppositely charged and thus moving away from $A$ toward $C$ needs more energy to encounter this attraction. In the middle, the energy is maximal being 0.8~eV. It decreases again because the attractive term increases again until the defect reaches site $C$, which is equivalent to site $A$ (considering the periodic boundary conditions). Remarkably, going from site $A$ to $D$ shows similar energy barrier to the $A$-$C$ path. This indicates that the Ti atom can also leave its \textit{pseudocube} and move into the neighboring one where the acceptor $V_{\mathrm{Sr}}$ is located [Fig. 4a]. In this $A$-$D$ switching, the four neighboring Ti atoms move all away to allow the Ti$_{\mathrm{Sr}}$ to migrate into the neighboring \textit{pseudocube}. This can be seen as a local defect diffusion.

In the experiment, switching the polarization is achieved by applying an electric voltage. This implies that the electron chemical potential (Fermi level) increases and according to our results this allows the compensated defect complex to capture two electrons, leading to a charged (Ti$_{\mathrm{Sr}}$+$V_{\mathrm{Sr}}$)$^{-2}$. This gives rise to two mid-gap states with 2$\mu_{B}$ as shown in Fig. 3. Overall, compared to the compensated complex, this charged counterpart shows similar barriers for switching [Fig. 4]. Noticeable decrease of the barrier along the $A$-$D$ path ($\sim$0.2~eV less) can be attributed to the fact that in the charged defect the Ti atom has more bounds with the neighboring oxygen atoms and shares the $t_{2g}$-state with the four neighboring Ti atoms [Fig. 3]. This is preserved upon moving to position $D$ and thus it eases the switching between $A$ and $D$ sites, compared to compensated (Ti$_{\mathrm{Sr}}$+$V_{\mathrm{Sr}}$) complex. It can also be the reason why the barrier along the $A$-$B$ decreases to about 70 meV. We note that along all paths, the intermediate \textit{images} of this charged defect complex preserve the localized mid-gap states and the magnetic moment of 2$\mu_{B}$. This preserve the semiconducting-like character.

For both compensated and charged defect complexes, the energy as well as the polarization magnitude in $A$, $B$, $C$, $D$ sites is equivalent. In real samples, switching between them, in a collective way or along different pathways, allows switching not only the polarization orientation, but also the defects-state distribution in real space [Fig. 4]. On the one hand, controlling the polarization direction is known to tune the potential barrier at interfaces, and thus the charge depletion/accumulation~\cite{agg+21npj}. On the other hand, the conductivity in defected semi-insulating systems, that is governed by charge hopping through defect states, can be strongly tuned upon switching because of the charge redistribution. Therefore, new behaviors such as $RS$ can be induced.

\textit{Conclusions}

In summary, we presented the potential of defects in intrinsically non-ferroelectric perovskites with SrTiO$_{3}$ as a prototype. Our calculations show that deep electron-donor, shallow electron-acceptor defects as well as compensating defect complex can be formed. The latter allows for pinning the Fermi level far from the gap edges. Increasing the Fermi level, the compensated defect complex can capture two electrons into the deep center component, giving rise to localized mid-gap states with a well-defined magnetic moment. The defect complex induces a spontaneous polarization of about 32 $\mu C/cm^{2}$. Allowed by STO symmetry, the defect can switch between the equivalent sites. Interestingly, switching the defect position switches both the polarization direction and the spatial distribution of the localized defect states. Depending on the switching pathway, energy barriers ranging between 0.1 and 0.8~eV are found. This is expected to enrich the switching possibilities, \textit{e.g.} upon applied voltages, and thus the material's response. Overall, with the advances in growth techniques with precise control of bulk/film stoichiometry, our results demonstrate the potential for achieving new behaviors by introducing and controlling defects.


\section*{Data availability}
Input and output files can be downloaded free of charge from the NOMAD Repository~\cite{drax-sche19jpm} at the following link: \url{https://doi.org/10.17172/NOMAD/2025.05.23-1}.

\section*{Acknowledgment} 
This work was supported by the project MeMaDe funded by the Leibniz Association under the Collaborative Excellence Program. W. A. acknowledge the funding by the Deutsche Forschungsgemeinschaft (DFG, German Research Foundation) - Projektnummer(510995508). The authors gratefully acknowledge the Gauss Centre for Supercomputing e.V. (\href{www.gauss-centre.eu}{www.gauss-centre.eu}) for funding this project by providing computing time on the GCS Supercomputer SuperMUC at Leibniz Supercomputing Centre (\href{www.lrz.de}{www.lrz.de}). The authors gratefully acknowledge the computing time granted by the Resource Allocation Board and provided on the supercomputer Lise and Emmy at NHR$\MVAt$ZIB and NHR$\MVAt$G{\"o}ttingen as part of the NHR infrastructure. The calculations for this research were conducted with computing resources under the project <bep00108>. W. A. thanks Martin Albrecht, Andreas Fiedler (Leibniz-Institut f\"{u}r Kristallz\"{u}chtung, Berlin), and Se{\'a}n R. Kavanagh (University College London) for fruitful discussions. 

\section*{Additional information}
Supplementary information is available for this paper at (/https:/...).

\newpage

\renewcommand{\thepage}{S\arabic{page}} 
\renewcommand{\thesection}{S\arabic{section}}  
\renewcommand{\thetable}{S\arabic{table}}  
\renewcommand{\thefigure}{S\arabic{figure}} 
\setcounter{figure}{0}
\twocolumngrid
\newpage
{\centering
{\large   
\textbf{Supporting Information: "Switchable polarization in non-ferroelectric SrTiO$_{3}$"}}}\newline

{\centering Wahib Aggoune$^{1,2}$ and Matthias Scheffler$^{1}$}\\

\noindent \textit{$^{1}$The NOMAD Laboratory at Fritz-Haber-Institut der Max-Planck-Gesellschaft, Faradayweg 4-6, 14195, Berlin, Germany}\\
\textit{$^{2}$Institut f\"ur Physik and CSMB, Humboldt-Universit\"at zu Berlin, Berlin, Germany}\\

\subsection*{Role of the XC functional and spin polarization}

To analyze the impact of the XC treatment on the defect characteristics, in Fig. S1 we show the Ti$_{\mathrm{Sr}}$ geometry optimized either with PBEsol or HSE(20) as well as with or without spin polarization. From first view, all the geometries look similar in which the Ti$_{\mathrm{Sr}}$ moves along the [011] direction. For example, the geometries optimized with PBEsol including/ignoring spin polarization are almost similar [Fig. S1a,c]. However, comparing the spin-polarized structures given by PBEsol and HSE(20) [Fig. S1c,d], one can see, with careful inspection, that the geometries differ in the arrangement of the oxygen atoms near the Ti$_{\mathrm{Sr}}$. The latter reflects the hybridization of the defect's orbitals (Ti-$d$) with those of the neighboring oxygen atoms, which is better described with HSE(20). This strongly affects the band structure as we will see in the following. Ignoring now the spin polarization in the HSE(20) relaxation [Fig. S1b], we see that the Ti$_{\mathrm{Sr}}$ defect moves along the [111] instead of [011] direction.

The resulting band structures in Fig. S1 are all obtained using the HSE(20), but either with or without spin polarization. We clearly see that in the case where the geometry is optimized using PBEsol, the two electrons given by Ti$_{\mathrm{Sr}}$ defect (donor) are hosted by the CBM [Fig. S1a], despite the band structure being computed with HSE(20). The charge is delocalized in the real space over all Ti atoms. Relaxation with PBEsol, including spin-polarized, gives one localized mid-gap state occupied by one electron. The second electron remains at the CBM, being delocalized in real space [Fig. S1c]. 

\begin{figure}[h!]
 \begin{center}
\includegraphics[width=.49\textwidth]{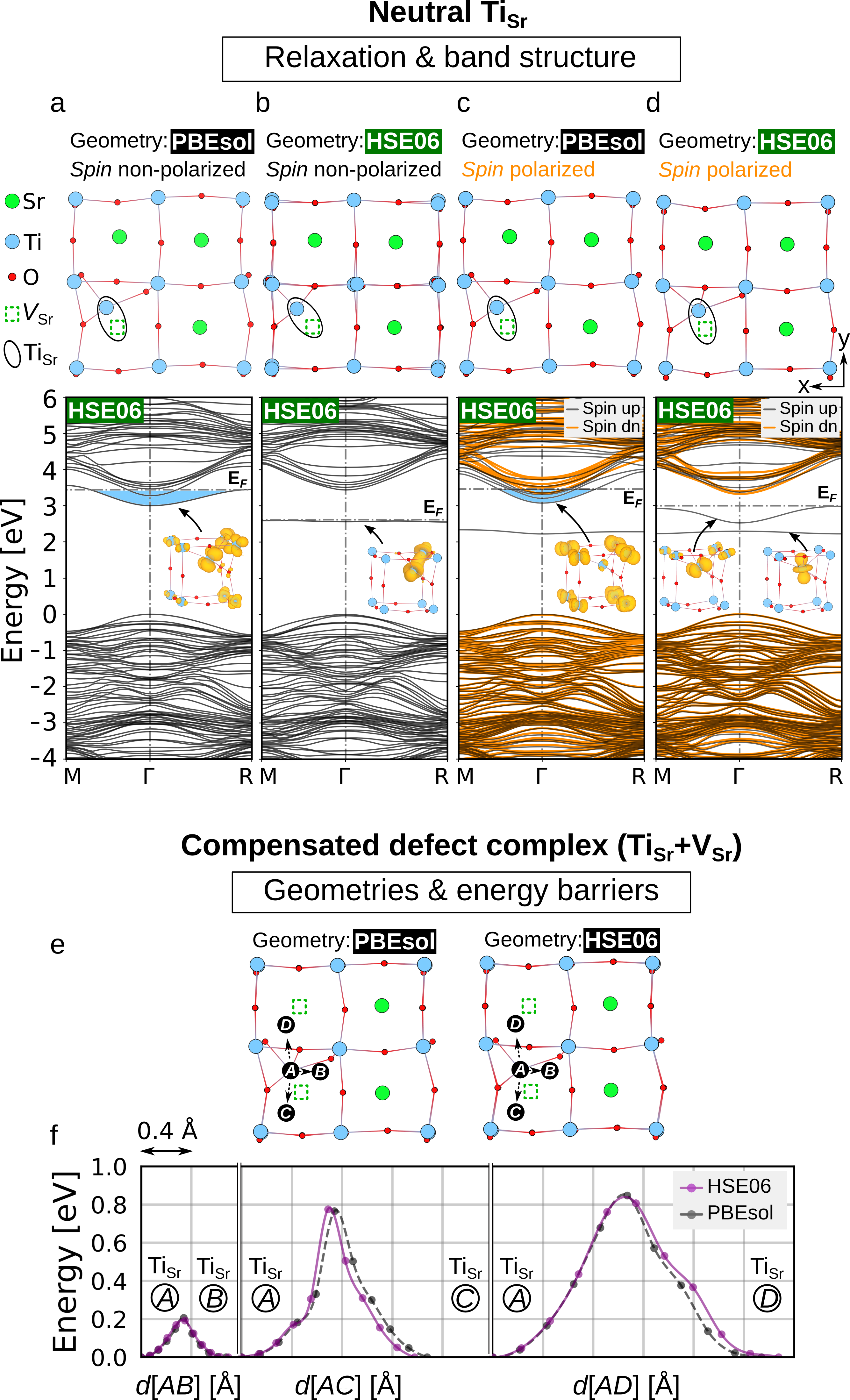}%
\caption{(Top) Impact of the XC functional on the relaxation and the band structure of neutral Ti$_{\mathrm{Sr}}$. All band structures are computed with HSE06($\alpha=$20$\%$), while the geometries are relaxed with PBEsol in (a) and (c) and HSE(20) in (b) and (d). The structures and band structures in (c) and (d) are computed with spin-polarized. The blue colored area in (a) and (c) indicates the occupied CBM. The bulk-like VBM is set to zero, while the Fermi level is indicated with a dashed vertical line. (Bottom) Impact of the XC functional on the structure and the energy barrier for switching the compensated defect complex (Ti$_{\mathrm{Sr}}$+$V_{\mathrm{Sr}}$). (e) The sites $A, B, C, D$ are highlighted on the geometries. (f) The energy barriers for switching the defect between the equivalent sites, computed with PBEsol and HSE(20) using their relaxed geometries. The \textit{x}-axis is the distance \textit{d} between the \textit{initial} and \textit{final} Ti$_{\mathrm{Sr}}$ positions along each path.}
\label{fig:bulk}
 \end{center}
\end{figure}

Only relaxation with HSE(20) along with spin-polarized allows us to correctly describe and predict the localized defect state. This configuration exhibits a magnetic moment of 2$\mu_{B}$. The two electrons are hosted by two localized mid-gap states. The lowest- (highest)-energy band exhibits $e_{g}$-like ($t_{2g}$-like) symmetry [Fig. S1d]. Ignoring spin in both relaxation and band structure of HSE(20) affects the geometry, as discussed above, as well as the resulting band structure [Fig. S1b]. A localized mid-gap band is found which is occupied by two electrons [Fig. S1b]. This state is distributed over two Ti atoms as the Ti$_{\mathrm{Sr}}$ moves toward the corner [111]. The latter is +0.25~eV ($\sim 30~meV/f.u.$) higher in energy compared to the geometry optimized with HSE(20) including spin polarization. Overall, despite the band structure being computed with HSE(20), relaxation with PBEsol of such a defect drastically affects the electronic band structure. Therefore, HSE(20) with spin-polarized (or at least Hubbard functional) is required for the structural optimization of the neutral Ti$_{\mathrm{Sr}}$ defect. This conclusion is also valid for the charged (Ti$_{\mathrm{Sr}}$+$V_{\mathrm{Sr}}$)$^{-2}$ defect complex as it exhibits localized mid-gap states [see Fig. 3 in the main text].

In the peculiar case of compensated defect complex (Ti$_{\mathrm{Sr}}$+$V_{\mathrm{Sr}}$), the system has a semiconducting character without mid-gap defect states [Fig. 3 in the main text]. Consequently, PBEsol is found to be enough for the structural optimization. As depicted in Fig. S1e, both PBEsol and HSE(20) structures are almost identical. This structure is non-magnetic. The energy barriers for switching the defect position in this complex are also well captured with PBEsol [Fig. S1f]. Therefore, PBEsol could be enough to also optimize and explore the energy barriers of similar compensated defects in other titanates-based perovskites.

\subsection*{Polarization}
In Fig. S2, we show how the polarization value is estimated for the compensated defect complex (Ti$_{\mathrm{Sr}}$+$V_{\mathrm{Sr}}$). Our bulk reference is the AFD structure with $P1$ symmetry, which has a polarization. Therefore, we need to extract the latter polarization to properly estimate that induced by the defect. Plotting the polarization branch from the centrosymmetric (containing the defect) to the AFD (containing the defect) to a fully relaxed defect is the key for estimating the polarization magnitude and its direction. The approach used is discussed in section \textit{Method} of the main text. For all the structures in Fig. S2, we fix the lattice parameters to those of the fully relaxed defect (G3). The fully relaxed geometry is given by HSE(20). Because of the compensated charge, this defect complex is semiconducting without mid-gap states, as discussed above. Thus, PBEsol is used to estimate the polarization of the intermediate \textit{images} as it gives the same polarization value as HSE(20) [see the red and green circles in Fig. S2]. This is only valid for this peculiar case. For the neutral Ti$_{\mathrm{Sr}}$ defect and the charged (Ti$_{\mathrm{Sr}}$+$V_{\mathrm{Sr}}$)$^{-2}$ defect complex, HSE(20) is required to estimate the polarization because of the localized mid-gap states, which are not captured by PBEsol.

Looking now into the polarization curves, we see that the computed polarization values of the linearly-interpolated \textit{images} can be within different polarization branches [dashed red line in Fig. S2]. This is due to the change in the ions' order in a specific direction, compared to the reference. The difference between the equivalent branches represents the polarization quantum. Additional \textit{images} are considered once the polarization value moves between branches in order to properly estimate the polarization quantum. Mapping all points gives the full polarization branch. This allows us to determine the polarization induced by the defect referenced to the AFD $P1$ phase [black arrows in Fig. S2]. Overall, we found that the polarization induced by the compensated defect complex (Ti$_{\mathrm{Sr}}$+$V_{\mathrm{Sr}}$) is about 32 $\mu C/cm^{2}$, oriented along the Ti-off-center direction [110].

\begin{figure*}[h!]
 \begin{center}
\includegraphics[width=.55\textwidth]{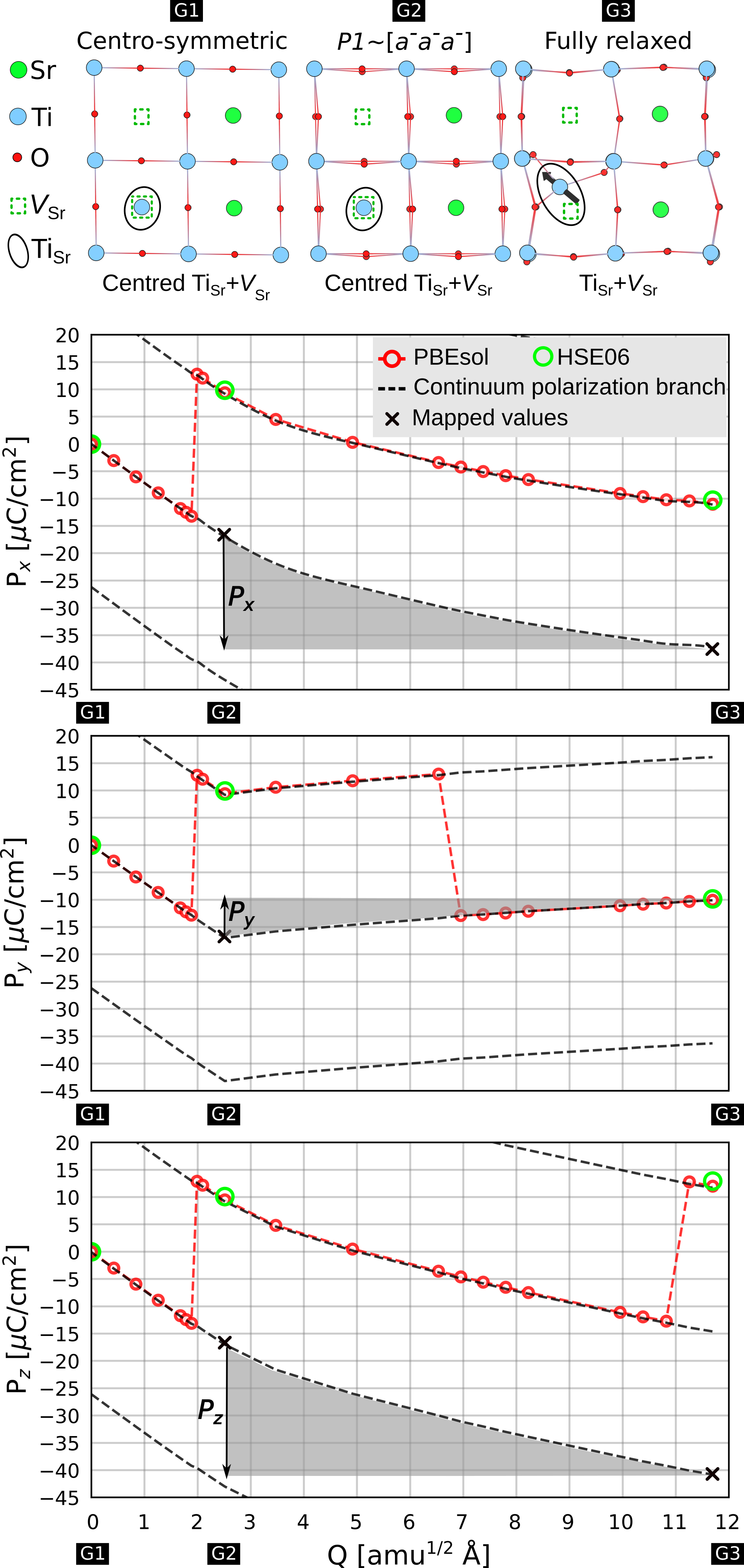}%
\caption{(Top) Sketch of the compensated defect complex (Ti$_{\mathrm{Sr}}$+$V_{\mathrm{Sr}}$). The configuration (G1) incorporates the defect in a centrosymmetric structure without relaxation. In the geometry (G2), the defect is incorporated into the $P1$ [$a^{-}a^{-}a^{-}$] structure without relaxation. In (G3), the defect is fully relaxed with HSE06($\alpha=$20$\%$). The calculated ($P_{x},~P_{y},~P_{z}$) components of spontaneous polarization for the structures going from geometry (G1) to (G2) to (G3) are shown in the (Bottom) panels. The intermediate \textit{images} between (G1), (G2) and (G3) are obtained with linear interpolation. The corresponding continuum polarization branches for each direction are highlighted with dashed black lines. The induced polarization by the fully relaxed defect is extracted and highlighted with black arrows by subtracting the polarization of geometry (G2) which is our bulk-like ($P1$) reference. All structures are given by HSE(20). The polarization values (red circles) are computed with PBEsol and the green ones are given by HSE(20), for the same HSE(20) geometries. The \textit{x}-axis in all plots is the mass-weighted configurational coordinate (Q) of each \textit{image}, referenced to the perfect structure (Q=0) as discussed in the \textit{Method} section of the main text.}

\label{fig:cell-def}
 \end{center}
\end{figure*}

\end{document}